\documentclass[10pt]{bmc_article}

\usepackage{cite} % Make references as [1-4], not [1,2,3,4]
\usepackage{url}  % Formatting web addresses  
\usepackage{ifthen}  % Conditional 
\usepackage{multicol}   %Columns
\usepackage{graphicx} 
 
\newboolean{publ}

\newenvironment{bmcformat}{\begin{raggedright}\baselineskip20pt\sloppy\setboolean{publ}{false}}{\end{raggedright}\baselineskip20pt\sloppy}

\begin{document}
\begin{bmcformat}


\title{Nature of protein family signatures: Insights from singular value analysis of position-specific scoring matrices}
 
%%%%%%%%%%%%%%%%%%%%%%%%%%%%%%%%%%%%%%%%%%%%%%
%%                                          %%
%% Enter the authors here                   %%
%%                                          %%
%% Ensure \and is entered between all but   %%
%% the last two authors. This will be       %%
%% replaced by a comma in the final article %%
%%                                          %%
%% Ensure there are no trailing spaces at   %% 
%% the ends of the lines                    %%     	
%%                                          %%
%%%%%%%%%%%%%%%%%%%%%%%%%%%%%%%%%%%%%%%%%%%%%%

\author{Akira R Kinjo\correspondingauthor$^{1}$%
       \email{Akira R. Kinjo\correspondingauthor - akinjo@protein.osaka-u.ac.jp}%
      \and
         Haruki Nakamura$^1$%
         \email{Haruki Nakamura - harukin@protein.osaka-u.ac.jp}
      }

%%%%%%%%%%%%%%%%%%%%%%%%%%%%%%%%%%%%%%%%%%%%%%
%%                                          %%
%% Enter the authors' addresses here        %%
%%                                          %%
%%%%%%%%%%%%%%%%%%%%%%%%%%%%%%%%%%%%%%%%%%%%%%

\address{%
  \iid(1)Institute for Protein Research, Osaka University,%
  Suita, Osaka 565-0871, Japan%
}%

\maketitle

%%%%%%%%%%%%%%%%%%%%%%%%%%%%%%%%%%%%%%%%%%%%%%
%%                                          %%
%% The Abstract begins here                 %%
%%                                          %%
%% The Section headings here are those for  %%
%% a Research article submitted to a        %%
%% BMC-Series journal.                      %%  
%%                                          %%
%% If your article is not of this type,     %%
%% then refer to the Instructions for       %%
%% authors on http://www.biomedcentral.com  %%
%% and change the section headings          %%
%% accordingly.                             %%   
%%                                          %%
%%%%%%%%%%%%%%%%%%%%%%%%%%%%%%%%%%%%%%%%%%%%%%

\begin{abstract}
        % Do not use inserted blank lines (ie \\) until main body of text.

        \paragraph*{Background:} 
Position-specific scoring matrices (PSSMs) are useful for detecting weak 
homology in protein sequence analysis, and they are
thought to contain some essential signatures of the protein families.
        \paragraph*{Results:} 
In order to elucidate what kind of ingredients constitute such family-specific 
signatures, we apply singular value decomposition to a set of PSSMs and  
examine the properties of dominant right and left singular vectors. 
The first right singular vectors were correlated with various 
amino acid indices including relative mutability, amino acid composition 
in protein interior, hydropathy, or turn propensity, depending on proteins. 
A significant correlation between the first left singular vector and 
a measure of site conservation was observed. It is shown that the contribution 
of the first singular component to the PSSMs act to disfavor potentially but 
falsely functionally important residues at conserved sites. 
The second right singular vectors were highly correlated with hydrophobicity 
scales, and the corresponding left singular vectors with contact numbers of 
protein structures. 
        \paragraph*{Conclusions:} 
It is suggested that sequence alignment with a PSSM is 
essentially equivalent to threading supplemented with functional information.
The presented method may be used to separate functionally important sites from 
structurally important ones, and thus it may be a useful tool for predicting 
protein functions.
\end{abstract}

\ifthenelse{\boolean{publ}}{\begin{multicols}{2}}{}

%%%%%%%%%%%%%%%%%%%%%%%%%%%%%%%%%%%%%%%%%%%%%%
%%                                          %%
%% The Main Body begins here                %%
%%                                          %%
%% The Section headings here are those for  %%
%% a Research article submitted to a        %%
%% BMC-Series journal.                      %%  
%%                                          %%
%% If your article is not of this type,     %%
%% then refer to the instructions for       %%
%% authors on:                              %%
%% http://www.biomedcentral.com/info/authors%%
%% and change the section headings          %%
%% accordingly.                             %% 
%%                                          %%
%% See the Results and Discussion section   %%
%% for details on how to create sub-sections%%
%%                                          %%
%% use \cite{...} to cite references        %%
%%  \cite{koon} and                         %%
%%  \cite{oreg,khar,zvai,xjon,schn,pond}    %%
%%  \nocite{smith,marg,hunn,advi,koha,mouse}%%
%%                                          %%
%%%%%%%%%%%%%%%%%%%%%%%%%%%%%%%%%%%%%%%%%%%%%%

%%%%%%%%%%%%%%%%
%% Background %%
%%
\section*{Background}

Protein sequence alignment using a position-specific scoring matrix (PSSM) or 
sequence profile \cite{Taylor1986,GribskovETAL1987} is 
now a standard tool for sequence analysis\cite{Eddy1996,AltschulETAL1997}.
Using a PSSM, it is often possible to detect very distantly related proteins 
which cannot be detected by the standard pairwise alignment based on a
position-independent amino acid substitution matrix (AASM).

An AASM is a 20$\times$20 real (usually symmetric) matrix each element of 
which reflects the tendency of substitution between amino acid residues. 
There have been many kinds of AASMs developed to date among which the most 
popular ones include the PAM \cite{DayhoffETAL1978} and the BLOSUM series \cite{BLOSUM}.
General properties of AASMs are now well clarified\cite{Altschul1991,TomiiANDKanehisa1996,KinjoANDNishikawa2004,PokarowskiETAL2007}.
Tomii and Kanehisa found that the PAM matrices can be well approximated by 
the volume and hydrophobicity of amino acid residues\cite{TomiiANDKanehisa1996}.
A similar result was obtained by Pokarowski et al.\cite{PokarowskiETAL2007},
but they also pointed out the importance of the coil preferences of amino acids 
residues.
Using eigenvalue decomposition, Kinjo and Nishikawa\cite{KinjoANDNishikawa2004}
showed that the most 
dominant component of AASMs is the relative mutability\cite{DayhoffETAL1978}
for closely related homologs, but it changes to hydrophobicity below 
the sequence identity of 30\%, and this transition
of dominant modes was related to the so-called twilight zone of sequence 
comparison\cite{Doolittle1986,Rost1999}. There are also AASMs specifically 
optimized to overcome the twilight zone \cite{KannETAL2000,QianANDGoldstein2002}.

Detection of very distant homologs is often possible by using PSSM-based sequence 
alignment methods such as PSI-BLAST\cite{AltschulETAL1997} or hidden Markov 
models\cite{Eddy1996,DurbinETAL} because a PSSM is specific to a particular 
protein family so that some family-specific features can be exploited.
In a PSSM, family-specific features are expressed as position-dependent 
substitution scores, and hence a PSSM is an $N\times$20 matrix where $N$ is 
the length of the protein or protein family it represents. 
Since PSSMs can be regarded as an extension of sequence 
motifs\cite{DurbinETAL}, family-specific features are, to the first 
approximation, a pattern of amino acid residues around functionally or 
structurally important sites expressed in a probabilistic manner.
In order to further understand the mechanism by which the effectiveness of 
PSSMs is realized, however, it is necessary to elucidate more general 
characteristics of PSSMs that are shared across different protein families.

To delineate the general properties of PSSMs, we analyze them by using 
singular value decomposition (SVD, Eq. \ref{eq:svd} in the Methods section).
By applying SVD, a PSSM can be decomposed into 20 orthogonal components of 
varying importance. Each singular component consists of a singular value 
(a scalar), right singular vector (r-SV) and left singular vector (l-SV).
See the Methods section for the details.
A singular value represents the relative importance of the component whereas
the corresponding r-SV (a 20-vector) represents a property of 20 amino acid 
types and the l-SV may be regarded as a one-dimensional (1D) numerical 
representation of the amino 
acid sequence that is ``dual'' to the property represented by the r-SV.
Since r-SVs can be regarded as amino acid indices\cite{KideraETAL1985a,NakaiETAL1988,TomiiANDKanehisa1996}, we can infer their meaning 
by comparing them with the entries of the AAindex database\cite{AAindex} which 
compiles many amino acid indices published to date.
This is a natural generalization of a previous work
where AASMs were analyzed by using eigenvalue decomposition \cite{KinjoANDNishikawa2004}.
The present analysis revealed a tendency of PSSMs that is analogous to the 
AASMs for close homologs. That is, the first principal component disfavors 
any substitutions and potentially functionally important residues are more 
severely penalized, and the second component is highly correlated with 
sequence and structural properties related to hydrophobicity. 
These features are expected to contribute to the effectiveness of 
sequence alignment based on PSSMs.

%%%%%%%%%%%%%%%%%%%%%%%%%%%%
%% Results and Discussion %%
%%
\section*{Results}

\subsection*{Overview}
\label{sec:overview}
In order to check to what extent a subset of singular components can explain 
the original PSSM, we calculated the accumulative contribution ratio
of each PSSM. The accumulative contribution ratio up to $k$-th singular 
value is defined as
\begin{equation}
  \label{eq:acr}
  S_k  = \frac{\sum_{\alpha=1}^{k}\sigma_{\alpha}}{\sum_{\alpha=1}^{20}\sigma_{\alpha}}
\end{equation}
where $\sigma_\alpha$ is the $\alpha$-th singular value which is non-negative.
The averages of $S_{k}$ for $k = 1, \cdots, 20$ are shown in 
Fig. 2. We observe that the first singular value 
contributes 17\% of the total singular values in the PDB set, and 24\% 
in the Pfam set. Thus, the contribution of the first singular component is 
relatively larger in the Pfam PSSMs than in the PSI-BLAST-generated PSSMs of 
PDB entries. This tendency may be related to the higher specificity of 
the Pfam hidden Markov models.
50\% contributions are made by first 4 or 5 components in the PDB or Pfam 
sets, respectively, whereas 90\% contributions are made by the first 15 
components in the both sets.
Compared to the case with AASMs where 50\% and 90\% contributions are made by 
first 3 and 10 singular values (or eigenvalues) \cite{KinjoANDNishikawa2004}, 
the ``compressibility'' of PSSMs is lower in the sense that more components 
are needed to explain the same fraction (50\% or 90\%) of the total components.
This is a reasonable result since each PSSM should contain some 
detailed information specific to the family to which the protein sequence 
belongs, whereas AASMs should contain more general information regarding 
the patterns of amino acid substitutions shared by many protein families.

In order to glance at the overall characteristics of decomposed PSSMs, we 
constructed 
a partial matrix $M_k$ for each PSSM by summing the first $k$ components, 
that is,
\begin{equation}
  \label{eq:pmat}
  M_{k} = \sum_{\alpha=1}^{k}\sigma_{\alpha}\mathbf{u}_{\alpha}\mathbf{v}_{\alpha}^{T}
\end{equation}
where $\mathbf{u}_{\alpha}$ and $\mathbf{v}_{\alpha}$ are the $\alpha$-th
left and right singular vectors, respectively,
and calculated the fraction of positive elements ($M_{20}$ is identical to the 
original PSSM). In both the PDB and Pfam sets, there are usually more negative 
elements than positive ones (Fig. 3). 
This is an expected behavior for log-odds matrices \cite{Altschul1991}. 
However, this skewed distribution is greatly pronounced for the $M_{1}$ 
matrices. In fact, most substitutions are disfavored by the first 
singular component of a PSSM. A typical example is shown in 
Fig. 1C where the contribution of the first component 
(i.e., $\sigma_1\mathbf{u}_1\mathbf{v}_1^{T}$) is purely negative.
Compared to $M_1$, other partial matrices ($M_k$ with $k > 1$) have 
more positive elements. This indicates that positive values in the final PSSM
must original from $M_k$ with $k>1$.
\subsection*{Characteristics of first singular components}
\label{sec:rsv1}
In order to interpret the physicochemical or biochemical meaning of the first
r-SVs ($\mathbf{v}_1$ in Eq.\ref{eq:svd}), we scanned the AAindex database and identified amino acid indices that 
frequently show significant correlations (Table 2).

In the PDB representative set, the most frequently correlated index was the 
relative mutability\cite{JonesETAL1992b} (AAindex: JOND920102) which is also 
the fifth most frequent index for the Pfam set. 
The relative mutabilities\cite{DayhoffETAL1978} represent the tendency of 
amino acid residues to be mutated during molecular evolution, and are not 
highly correlated with any other indices \cite{TomiiANDKanehisa1996}. 
It is thus expected that some intrinsic characteristics of protein evolution 
is embedded in their values. The relative mutability is the most dominant 
component in the ordinary (position-independent) AASMs targeted at closely 
related proteins\cite{KinjoANDNishikawa2004}.
As in the case of AASMs, the first r-SVs are negatively correlated with 
the relative mutability (Recall that all the elements of the first r-SVs are 
of the same sign in most cases so that we can make them all positive without 
losing generality). An example is shown in Fig. 4A.
Thus, noting that the first singular components (i.e., partial matrix 
$M_1$ in Fig. 3) are mostly negative, we can see that 
substitutions of those residues with low mutabilities are more severely 
penalized.

The interior composition of amino acids in intracellular proteins of 
mesophiles\cite{FukuchiANDNishikawa2001} (FUKS010106) is another 
frequently correlated index, ranked second and sixth in the PDB and Pfam sets, 
respectively. As we can see in the example shown in Fig. 4B, 
those residues that are less abundant in protein interior are more 
severely penalized. This seems to contradict our intuition that residues 
in the protein interior are more conservative than those on the protein 
surface. However, many functionally important residues exist on the surface
(ligand binding sites and catalytic sites, etc.).
Thus, these r-SVs should be regarded as representing potentially functionally 
important residues. Note, however, although these residues share some 
properties common to conserved residues, most of them are not actually 
important (otherwise they should not be penalized). 

Other frequently correlated indices shared by both PDB and Pfam sets are 
the conformational parameter of $\beta$-turn\cite{BeghinANDDirkx1975} 
(BEGF750103) and the hydropathy index of Kyte and 
Doolittle\cite{KyteANDDoolittle1982} (KYTJ820101).
The most frequently correlated index for the Pfam set was 
``principal component I'' of Sneath (SNEP660101) \cite{Sneath1966}. 
The name of this index is rather cryptic, but it is weakly negatively 
correlated with turn or coil propensities (data not shown).
These indices can be readily related to interior-surface propensities:
$\beta$-turns, coils, and hydrophilic residues tend to be on the surface
of a protein, and so on. The general trend is that substitutions of those 
residues that tend to be on the surface are more severely penalized
(Fig. 4C, D). Again, this may be due to the fact that many 
(potentially) functionally important residues are on the protein surface. 

It is noted that no single index is overwhelmingly dominant in the first r-SVs 
so that different PSSMs are characterized by different properties.
This is a reasonable result since each PSSM is specific to a particular 
protein family which is under the influence of specific evolutionary pressures
 and biological constraints. Nevertheless, relative mutability, 
hydrophobicity, and turn/coil propensity are the relatively more dominant 
characteristics of the first r-SVs.

If the first r-SV of a PSSM represents a property of amino acid residues that 
is well-conserved, then the first l-SV is expected to represent the pattern or 
extent of conservation of that property along the amino acid sequence.
One such measure is the information content (also referred to as 
Kullback-Leibler divergence or relative entropy \cite{CoverANDThomas}) 
which is a kind of distance of the distribution of amino acid residues at 
a given site of the sequence from the background distribution.
The information content $D_i$ of site $i$ is defined as 
\begin{equation}
  \label{eq:divergence}
  D_i = \sum_{a}P_i(a)\log [P_i(a)/Q(a)]
\end{equation}
where $P_i(a)$ is the frequency of amino acid type $a$ at the site $i$
and $Q(a)$ is the background frequency of amino acid type $a$.
In general, information content tends to be larger at more conserved sites.
This information is available in the PSSMs created with PSI-BLAST. 
A significant correlation was found between the first l-SVs 
and information content of PSSMs of the PDB set 
with correlation coefficient of 0.543 on average ($P < 10^{-17}$, assuming 
the average sequence length of 217 residues). The median of the correlation 
coefficient was 0.601 indicating that the correlation is even higher for many 
of the PSSMs. When calculating the correlation coefficient, we converted the 
signs of the elements of the l-SV so that most elements become positive.
Thus, a positive correlation implies that a site with a large value of 
the first l-SV element usually has high information content, indicating that 
substitutions at those sites with more information content are more 
severely penalized.
An example of such correlation is shown in Fig. 5.
l-SVs other than the first one did not show high correlations with 
information content (data not shown).
For those PSSMs whose first r-SVs are highly correlated with JOND920102 
(110 entries), 
FUKS010106 (74), BEGF750103 (56), KYTJ820101 (49), and SNEP660101 (24) (Table 2), 
the average correlation coefficients were 0.646, 0.703, 0.654, 0.536, and 0.593,
respectively. Thus, the high correlation between the first l-SV and information 
content is not limited to specific PSSMs whose first r-SVs are correlated to 
some particular indices.

% The contribution of the first singular component to the original 
% matrix is mostly negative (Fig. 3) while the first r-SVs may be  
% regarded as representing potentially functionally important residues, and 
% the first l-SVs are correlated with information content of conservation.
% Thus, the first singular component makes negative contributions even to 
% conserved residues, especially at conserved sites.
% As will be illustrated later by an example, the score of a conserved residue 
% is made positive
% by intricate contributions from several singular components (from the second 
% to twentieth components) in a coherent manner. 
% But these contributions may also make the scores of some other 
% non-conserved residues positive.
% Such residues are what we called 
% ``potentially functionally important residues'' above. 
% By the negative contribution from the first singular component, the scores 
% of all the residue types at all the sites are (generally) decreased, 
% but only those of conserved residues remain positive.

\subsection*{Characteristics of second singular components}
\label{sec:rsv2}
In the same manner as the first r-SVs, we searched for indices that are 
highly correlated with the second r-SVs of the PSSMs (Table 3).
In this case, relative partition energies derived by the Bethe approximation
\cite{MiyazawaANDJernigan1999} (AAindex: MIYS990101) is the most 
correlated index in 33\% of the PDB set and 54\% of the Pfam set.
This index is a kind of hydrophobicity scale.
Furthermore, other frequently correlated indices, such as 
interactivity scales of Bastolla et al. \cite{BastollaETAL2005} 
(BASU050101, BASU050103), polarity \cite{Grantham1974} (GRAR740102),
optimal matching hydrophobicity \cite{SweetANDEisenberg1983} (SWER830101),
and all other indices in Table 3, are all related to 
hydrophobicity scales. The ten most frequently correlated indices alone 
match 85\% and 94\% of the second r-SVs of the PSSMs in the PDB and Pfam sets, 
respectively.
Therefore, while the first r-SVs are of diverse characteristics, the second 
r-SVs are almost exclusively determined by hydrophobic properties.
It is interesting to note that the hydropathy index of Kyte and 
Doolittle \cite{KyteANDDoolittle1982} which was found to be correlated to some
first r-SVs (Table 2) was not found to be the the index most 
correlated with the second r-SVs in most cases. Although the hydropathy index
is highly correlated with the partition energy of Miyazawa and Jernigan 
(correlation coefficient of -0.84), there seems to be a meaningful difference
between them.

The correlation between the second r-SVs and hydrophobicity scales is striking.
Therefore, it is expected that the second left singular vectors (l-SVs) are 
correlated with some structural property that is dual to the hydrophobicity.
One such structural property is the contact number \cite{NishikawaANDOoi1980,NishikawaANDOoi1986,KinjoETAL2005}, 
which is the number of residues in contact with a given residue in a native 
protein structure.
We calculated contact numbers of the PDB set (based on the definition by Kinjo et al. \cite{KinjoETAL2005}) and their correlations with the 
second l-SVs. The average correlation coefficient was 0.511 (standard 
deviation 0.113) which is highly significant ($P < 10^{-15}$) for the 
average protein length of 217 residues in the PDB set.
Fig. 6 shows an example of the highly correlated second l-SV and 
contact numbers. 

Recall that the elements of the first r-SVs were of the same sign in most 
cases (Fig. 3). Thus, by the orthogonality of singular vectors, 
the elements of the second r-SVs should necessarily contain values of both 
signs in most cases. 
The same argument also applies to l-SVs.
Therefore, the contribution from the second component of a PSSM, namely 
$\sigma_{2}\mathbf{u}_{2}\mathbf{v}_2^{T}$, contains both positive and negative 
elements corresponding to favorable and unfavorable substitutions, respectively.
Now let $\mathbf{w}$ represent the relative partition energy of 
Miyazawa and Jernigan \cite{MiyazawaANDJernigan1999} (MIYS990101), and 
$\mathbf{n}$ represent the contact number vector of a protein standardized by 
subtracting the average value from each element.
We calculated the correlation coefficient between the two matrices $\mathbf{u}_2\mathbf{v}_2^{T}$ and $\mathbf{w}\mathbf{n}^{T}$ for those 361 proteins whose 
second r-SVs are most correlated with $\mathbf{w}$. We obtained the average 
correlation of -0.45 which is highly significant ($P < 10^{-220}$) 
taking into account the average number of elements (217 $\times$ 20). 
Since hydrophilic and hydrophobic residues have positive and negative 
partition energies, respectively, the negative correlation means that
hydrophobic residues with high contact numbers (buried)
and hydrophilic residues with low contact numbers (exposed) are more favored 
compared to hydrophobic residues with low contact numbers and hydrophilic 
residues with high contact numbers.
Thus, within the framework developed here, we can consider the second 
singular component represents the structural stability of the protein.

\subsection*{Characteristics of third and other singular components}
The indices that are most frequently correlated with the third r-SVs of 
the PSSMs are listed in Table 4.
In general, the third r-SVs are correlated with those indices related 
to the volume or bulkiness of amino acid residues such as 
CHAM830106, SNEP660103, LEVM760102, LEVM760105 and OOBM770105 
(see Table 1 for descriptions).
Another kind of index common to the PDB and Pfam sets is 
the $\alpha$-helix propensity derived from designed 
sequences\cite{KoehlANDLevitt1999b} (KOEP990101) which is actually correlated 
with coil propensity (data not shown). This index was also found to be 
frequently correlated with the fourth r-SVs.
A structural quantity that may be associated with bulkiness of amino acid 
residues is the volume of the ``territory'' of residues as defined by 
the Voronoi tessellation\cite{Richards1977,Qhull}. When we compared the Voronoi 
volumes calculated from protein structures with the third l-SV,
we observed a significant but weak correlation of 0.345 ($P < 0.0003$).
(The Voronoi volume of a residue was calculated by summing the Voronoi 
volumes of the atoms that belong to the residue; only half of the residues 
with smaller volumes are used for comparison as surface residues with 
[sometimes infinitely] large volumes are not meaningful.)
If we limit the comparison to those proteins whose third r-SVs are most 
correlated with CHAM830106 (214 entries), SNEP660103 (201) or LEVM760102 (137),
the correlations were 0.366, 0.251, or 0.479, respectively.
Therefore, the correlation of the third l-SV to the Voronoi volume is 
significant, but not 
as consistent as those of the first and second l-SVs to information content and
contact numbers, respectively.

The propensity of the fourth and fifth r-SVs are not so clearly characterized 
as the first three r-SVs, but helix (KOEP990101) and helix cap propensities \cite{AuroraANDRose1998} as well as some bulkiness parameters are relatively highly
correlated with the fourth r-SVs, while the net charge 
(KLEP840101) \cite{KleinETAL1984} and $\alpha$-NH chemical shifts (BUNA790101)
\cite{BundiANDWuthrich1979} were the indices most correlated with the fifth 
r-SVs of more than 30\% of the PSSMs in both the PDB and Pfam sets.

\subsection*{Example: Conserved sites in the globin family}
\label{sec:globin}
To illustrate the points made above, we now examine the PSI-BLAST PSSM of 
a globin (PDB 3sdhA\cite{3SDH}, hemoglobin I 
from \emph{Scapharca inaequivalvis}).
The globin family is one of the most extensively studied protein 
families\cite{BashfordETAL1987,LecomteETAL2005}. 
Ota et al.\cite{OtaETAL1997} examined in detail seven highly conserved 
residues in globins identified by Bashford et al.\cite{BashfordETAL1987} 
(namely, the sites B10, C2, CD1, CD4, E7, F4, and F8, according to the 
numbering scheme of Bashford et al.\cite{BashfordETAL1987}),
and succeeded in separating structurally important sites from functionally 
important sites.
Fig. 7 shows the contributions of various components to 
the seven highly conserved sites studied in Ota et al. \cite{OtaETAL1997}.
The most correlated amino acid indices for the first 5 r-SVs are 
BEGF750103, MIYS990101, FAUJ880106, KOEP990101, and AURR980119 
(see Table 1 for their descriptions).

The contributions to those conserved residues that were identified as 
functionally important by 
Ota et al. \cite{OtaETAL1997} (namely, E7 and F8) are mainly from the third
 and fifth components which are correlated to bulkiness and helix capping 
propensity, respectively. Other conserved residues were identified as 
structurally important, and their scores consist mainly of the second singular 
component which is related to the hydrophobicity, except for the proline 
residue at the C2 site to which the helix capping propensity is the main 
contributor.
These observations are consistent with the analysis of Ota et al. \cite{OtaETAL1997} which was based on three-dimensional profiles\cite{BowieETAL1991,OtaANDNishikawa1997}.

The contributions of the first singular component to these sites are all
negative for all residues (Fig. 7) which is consistent with 
the general argument provided above. 
We now consider the meaning of the negative contribution of the first 
singular component.
For simplicity, we first consider the site F8 where the histidine residue 
is perfectly conserved. At this site, only the score of histidine should be 
positive and all others be negative.
Positive contributions to the score of histidine is made from 
the third, fifth and other singular components so that the total contributions 
from second to twentieth components are as large as 14.
Without the contribution from the first singular component, the scores of 
some other residues such as asparagine and tyrosine are also positive 
although not as large as that of histidine.
Thus, we can see that the large positive score of a conserved residue 
(histidine) is made by coherent contributions from multiple singular components
whereas the scores of residues that are not conserved may be positive but 
small due to incoherent contributions. Nevertheless, positive scores of 
non-conserved residues degrades the specificity of a PSSM. Thus, they should 
be somehow made negative. Similar arguments apply to other conserved sites 
except that different residues may be conserved at different sites for 
different reasons.
The score of potentially but falsely functionally important residues at all 
conserved sites can be made negative at once by simply subtracting the scores 
according to the common properties of amino acid residues at these sites, 
and this is the role of the first singular component. 
In the present example, the common property happened to be related to the 
$\beta$-turn propensity.

\section*{Discussion}
    
Kinjo and Nishikawa\cite{KinjoANDNishikawa2004} analyzed a set of amino acid 
substitution matrices constructed from multiple alignments of protein families 
of varying percent sequence identities (\%ID). It was found that, at high 
\%IDs ($> 35$\%), the first and second most dominant components were 
correlated with relative mutability and hydrophobicity, respectively, 
while at low \%IDs ($< 30$\%), the order was opposite (hydrophobicity first, 
and then the relative mutability). It was suggested that the dominance of 
the relative mutability over hydrophobicity patterns is the prerequisite for 
reliable detection of homologs.
In the case of PSSMs, the characteristics of the first singular component 
may vary depending on the protein (family). Nevertheless, the first singular 
components seem to represent some functional constraints which disfavor any
substitutions, and the second (and third) singular components are 
predominantly determined by 
such structural requirements as hydrophobicity (and packing).
Although both functional and structural constraints are important for distant
homolog detection, the dominance of the former over the latter may be more 
influential for the high specificity of sequence alignment methods based 
on PSSMs. Noting again that the Pfam PSSMs have larger first singular values 
(Fig. 2) and their first components contain more
negative elements (Fig. 3) compared to PSI-BLAST-generated PSSMs of
the PDB set, 
this view of the first singular component is consistent with a general 
observation that Pfam PSSMs exhibit, on average, higher specificity than 
those generated by PSI-BLAST.

As pointed out by Tomii and Kanehisa \cite{TomiiANDKanehisa1996}, 
side-chain volume and hydrophobicity are the main ingredients of AASMs.
In addition to these two properties, Pokarowski et 
al.\cite{PokarowskiETAL2007} also noted the importance of the coil propensity. 
Wrabl and Grishin\cite{WrablANDGrishin2005} also found 
similar preferences in the study of properties extracted from multiple 
sequence alignments.
These properties are also found to be the main ingredients 
of PSSMs in the present study. Some of the first r-SVs showed significant 
correlation to indices related to coil propensity such as BEGF750103 and 
SNEP660101 (Table 2); hydrophobicity is predominant in the 
second r-SVs; and side-chain volumes often show high correlation with
the third r-SVs. 

Bastolla et al.\cite{BastollaETAL2005} have studied the correlation between
the ``interactivity'' scale of amino acid residues and the principal 
eigenvectors of the native contact maps\cite{PortoETAL2004}.
Their interactivity scale is a kind of hydrophobicity scale, obtained by
eigenvalue decomposition of a contact potential and subsequent optimizations.
The principal eigenvector of a contact map is known to contain almost 
sufficient information for recovering the native structure itself\cite{PortoETAL2004}, and is highly correlated with contact number vector \cite{KinjoANDNishikawa2005}.
Bastolla et al.\cite{BastollaETAL2005} showed that the interactivity scales 
aligned along the amino acid sequence of a protein, then averaged over homologs,
 were significantly correlated with the principal eigenvector with the average 
correlation coefficient of 0.47. Note that the interactivity scales of 
Bastolla et al. are found among those indices that are most correlated with 
the second r-SVs in Table 3 (BASU050101 and BASU050103), and that
the second l-SVs are correlated with contact number vectors.
Thus, the present result is not only consistent with that of Bastolla et al. \cite{BastollaETAL2005}, but also demonstrates that some structural information 
is already embedded in a PSSM, which also explains why contact numbers can be 
predicted at high accuracy by using PSSMs\cite{KinjoETAL2005,Yuan2005,KinjoANDNishikawa2005c,IshidaETAL2006,KinjoANDNishikawa2006}.

%%%%%%%%%%%%%%%%%%%%%%
\section*{Conclusions}
We analyzed PSSMs by decomposing them into singular components.
The characteristics of the first right singular vectors was found to vary 
depending on protein families, but the corresponding left singular vectors 
showed high correlation with information content. The  contributions of the 
first singular components to the original PSSMs are usually negative 
so that the substitutions of potentially but falsely functionally important 
residues at conserved sites are more severely penalized. 
The second right singular vectors were almost always related to hydrophobicity
of amino acid residues, and the left singular vectors are significantly 
correlated with contact number vectors, thus demonstrating that the structural
information is directly embedded in the PSSMs. Other structural information
seem to be also included in the PSSMs, although not as significantly as 
hydrophobicity and contact numbers. Therefore, sequence alignment using PSSMs
may be regarded as threading \cite{BowieETAL1990,BowieETAL1991,JonesETAL1992}
supplemented with some functional information.
Based on the present analysis, it may be possible to define \emph{a priori} 
measure of the quality of PSSMs which may lead to a rational strategy for 
constructing more effective PSSMs by mixing various functionally/structurally
relevant contributions with appropriate singular values.
Finally, the illustrated example (Fig. 7) suggests that the 
present methodology may be used for discerning functionally important sites
from structurally important ones, and hence be useful for the prediction of
protein functions.

%%%%%%%%%%%%%%%%%%
\section*{Methods}
\subsection*{Singular value decomposition of position-specific scoring matrix}
A position-specific scoring matrix (PSSM) is a real rectangular matrix of 
size $N\times 20$ where $N$ is the length of the amino acid sequence of 
a protein (or protein family). We assume $N > 20$ although this condition is 
not strictly necessary. 
Each column of a PSSM corresponds to an amino acid type, whereas each row
corresponds to a site in the amino acid sequence.
Let $M = (M_{ij})$ be a PSSM. The element $M_{ij}$ represents the score for 
the amino acid $j$ at the site $i$ (Fig. 1A). By applying 
singular value decomposition \cite{MatrixAnalysis}
(SVD), we have
\begin{equation}
  \label{eq:svd}
  M = U\Sigma V^{T} = \sum_{\alpha=1}^{20}\sigma_{\alpha}\mathbf{u}_{\alpha}\mathbf{v}_{\alpha}^{T}
\end{equation}
where $U = (\mathbf{u}_{1},\cdots,\mathbf{u}_{20})$ and 
$V=(\mathbf{v}_1, \cdots, \mathbf{v}_{20})$ are $N\times 20$ and 
$20\times 20$ orthogonal matrices, respectively, that is,
$\mathbf{u}_{\alpha}^{T}\mathbf{u}_{\beta} = \delta_{\alpha\beta}$ and 
$\mathbf{v}_{\alpha}^{T}\mathbf{v}_{\beta} = \delta_{\alpha\beta}$ 
($\delta_{\alpha\beta}$ is Kronecker's 
delta). An example of SVD of a PSSM is given in Fig. 1.
The 20-vectors $\mathbf{v}_{\alpha}$'s are called right singular vectors 
(r-SV, Fig. 1C).
Since each element of a right singular vector numerically represents 
some property of an amino acid type, we can regard a right singular 
vector of a PSSM as an amino acid index
\cite{KideraETAL1985a,NakaiETAL1988,TomiiANDKanehisa1996} (possibly 
specific to the parent PSSM).
The $N$-vectors $\mathbf{u}_\alpha$'s are called left singular vectors 
(l-SV, Fig. 1C).
Since each element of a left singular vector numerically represents 
some property of the corresponding site in the sequence, we can regard a left
singular vector of a PSSM as a generalized 1D structure.
$\Sigma = \mathrm{diag}(\sigma_{1}, \cdots, \sigma_{20})$ 
is a diagonal matrix whose elements are the singular values of the PSSM, 
sorted in the decreasing order (Fig. 1B). 
Singular values are always non-negative and 
their magnitudes represent relative importance of the corresponding 
singular components (i.e., the pair of right and left singular vectors).

\subsection*{Data sets}
We analyze two sets of PSSMs. One is a representative set derived from 
the Protein 
Data Bank (PDB)\cite{PDB} and the other is the Pfam database\cite{Pfam}.

The representative protein chains in the PDB were 
obtained from the PISCES server \cite{PISCES} with cutoffs of 25\% sequence identity, 20\% R-factor, 
2.0\AA{} resolution and sequence length ranging from 40 to 500. 
Only the structures determined by X-ray crystallography were used. 
Those proteins which were classified as all-$\alpha$, all-$\beta$, 
$\alpha/\beta$, $\alpha+\beta$, 
multi-domain, or small proteins according to the SCOP (version 1.71) 
\cite{SCOP} database were retained. As a result, we obtained 1096 protein 
chains.
For each of these proteins, a PSSM was created by running 
PSI-BLAST against the UniRef100 protein sequence database 
(release 12.1) \cite{UniRef} with e-value cutoff of 0.0005 and 3 iterations.

Although Pfam is a database of hidden Markov models of protein 
families\cite{DurbinETAL}, we can regard its entries as PSSMs by using only the 
scores for matching states.
We extracted from Pfam release 22.0 (July 2007) those proteins 
whose sequence lengths were at least 40 residues, resulting in 8869 protein 
families. 

\subsection*{Searching AAindex}
As mentioned above, each right singular vector (r-SV) can be regarded as an 
amino acid index, a set of numerical values reflecting some property of 
amino acid residues. In order to clarify the meaning of each
r-SV, we scanned the AAindex database \cite{TomiiANDKanehisa1996,AAindex} 
(Release 9.1, August, 2006) which compiles many amino acid 
indices published to date. 
For a given $\alpha$ (= 1, 2, $\cdots$, 20), the amino acid index that showed 
the highest correlation to the $\alpha$-th r-SV of each PSSM were identified.
If the absolute value of the correlation coefficient between the index and the 
r-SV is greater than or equal to 0.6, then the index is counted as significant.
Identified indices are sorted according to the number of times they are counted 
as significant.
In Table 1, we summarize the descriptions of the AAindex entries 
that will be mentioned in the Results section.
 
%%%%%%%%%%%%%%%%%%%%%%%%%%%%%%%%
\section*{Authors contributions}
ARK conceived of the study, wrote computer programs, collected and analyzed data,
and wrote the manuscript. HN revised the manuscript. Both authors approved the final 
manuscript.

%%%%%%%%%%%%%%%%%%%%%%%%%%%
\section*{Acknowledgements}
  \ifthenelse{\boolean{publ}}{\small}{}
The authors thank Daron M. Standley for critically reading the manuscript.
This work was supported in part by a grant-in-aid from the Japan Science and 
Technology Agency.

%%  Bmc_article.bst  will be used to                       %%
%%  create a .BBL file for submission, which includes      %%
%%  XML structured for BMC.                                %%

{\ifthenelse{\boolean{publ}}{\footnotesize}{\small}
 \bibliographystyle{bmc_article}  % Style BST file
  \bibliography{mypaper,refs} }     % Bibliography file (usually '*.bib' ) 

%%%%%%%%%%%

\ifthenelse{\boolean{publ}}{\end{multicols}}{}

%%%%%%%%%%%%%%%%%%%%%%%%%%%%%%%%%%%
%%                               %%
%% Figures                       %%
%%                               %%
%% NB: this is for captions and  %%
%% Titles. All graphics must be  %%
%% submitted separately and NOT  %%
%% included in the Tex document  %%
%%                               %%
%%%%%%%%%%%%%%%%%%%%%%%%%%%%%%%%%%%

%%
%% Do not use \listoffigures as most will included as separate files

\section*{Figures}
  \subsection*{Figure 1 - 
  Example of singular value decomposition of a PSSM (c.f. Eq. \ref{eq:svd}).}
A: The original PSSM (based on the PDB entry 3sdhA \cite{3SDH}); 
B: Singular values; 
C: Pairs of left singular vector (l-SV) $\mathbf{u}_\alpha$ of $N$ dimensions 
and right singular vector (r-SV) $\mathbf{v}_\alpha$ of 20 dimensions ($\alpha = 1, \cdots, 20$). The abscissa indicates residue number for the left singular vectors (l-SV), and amino acid type for the right singular vectors (r-SV).
The ordinate shows the vector elements relative to zero
(note that only the relative values, not absolute ones, are meaningful).
\begin{center}
  \includegraphics[width=16cm]{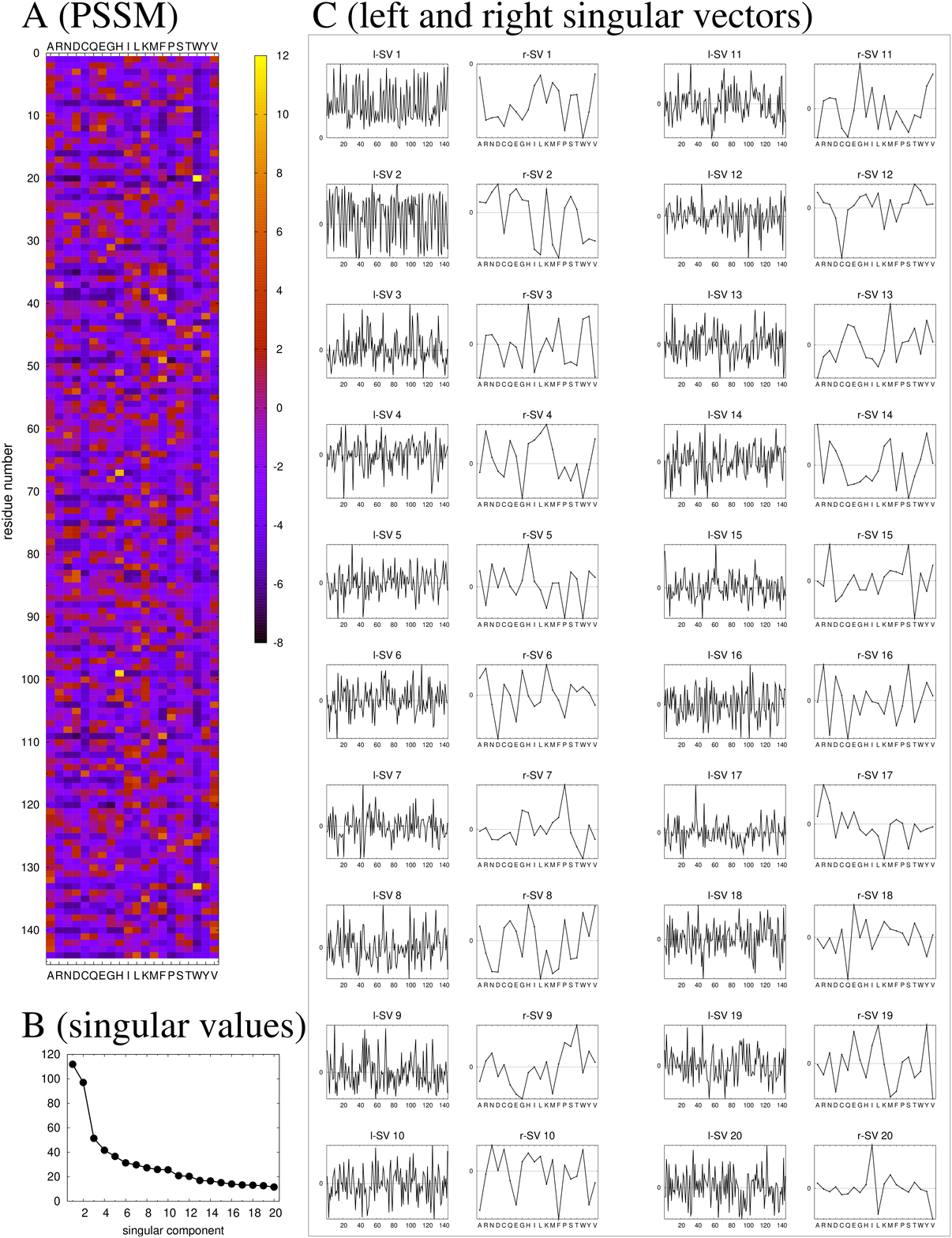}
\end{center}

  \subsection*{Figure 2 - Accumulative contribution ratio ($S_k$\%, $k = 1,\cdots, 20$) 
averaged over the PDB and Pfam sets.} 
\begin{center}
  \includegraphics[width=8cm]{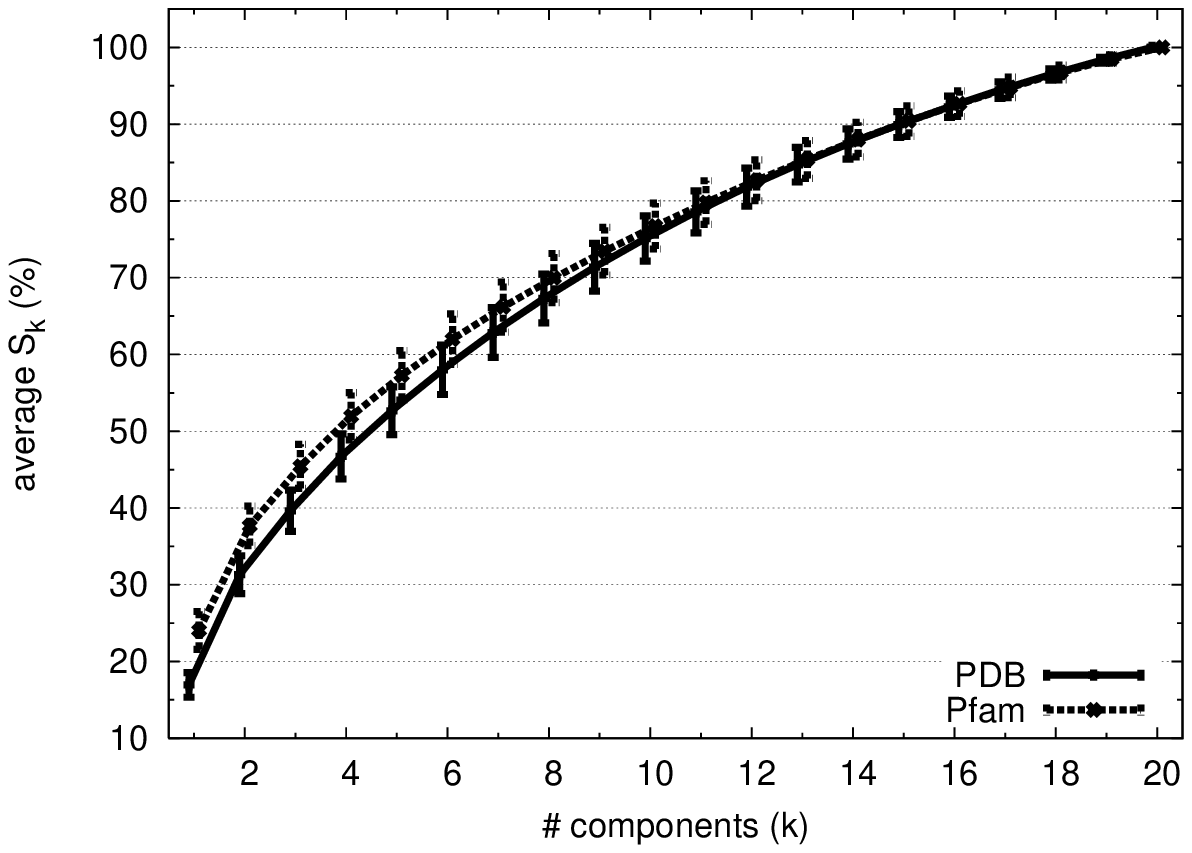}
\end{center}

  \subsection*{Figure 3 - Fraction of positive elements in partial matrices $M_{k}$ averaged over the PDB and Pfam sets.} 
\begin{center}
  \includegraphics[width=8cm]{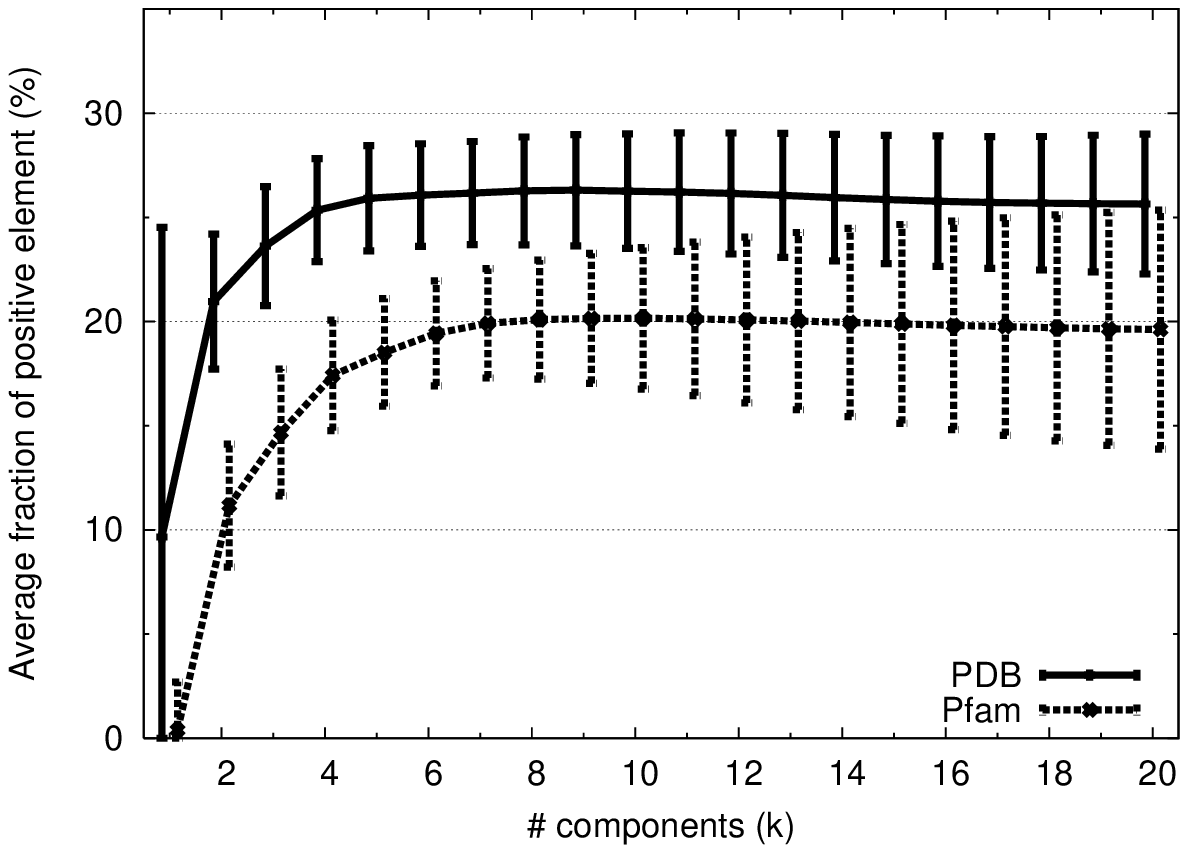}
\end{center}

  \subsection*{Figure 4 - Examples of correlations between the first r-SV and amino acid indices.}
The abscissa of each panel indicates the value of elements in the first 
right singular vector of a PSSM where
the signs are determined by making the value for cysteine positive.
The ordinates are (A) the relative mutability \cite{JonesETAL1992b}, 
(B) the interior composition of amino acids in intracellular proteins of 
mesophiles\cite{FukuchiANDNishikawa2001},
(C) the hydropathy scale\cite{KyteANDDoolittle1982}, and 
(D) the conformational parameter of beta-turn\cite{BeghinANDDirkx1975}.
The labels on the ordinates indicate the identifiers of the AAindex database
(Table 1).
\begin{center}
  \includegraphics[width=16cm]{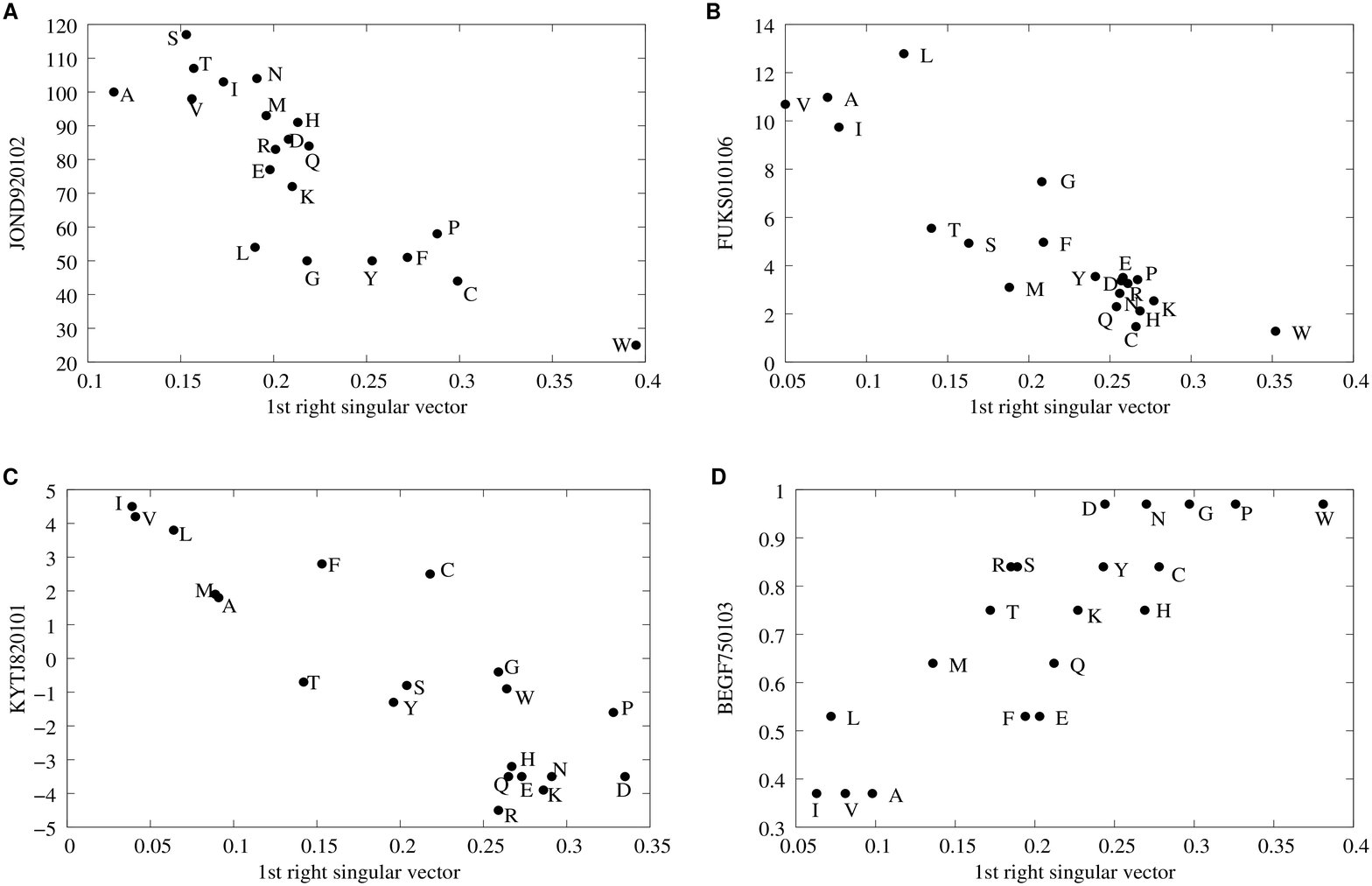}
\end{center}

  \subsection*{Figure 5 - Example of the first left singular vector and information content.}
Shown are the first l-SV and information content of the 
PDB entry 1e6uA\cite{1E6U}. The 
values of the l-SV elements are scaled by 20 times to match the information 
content. The correlation coefficient is 0.76.
\begin{center}
 \includegraphics[width=8cm]{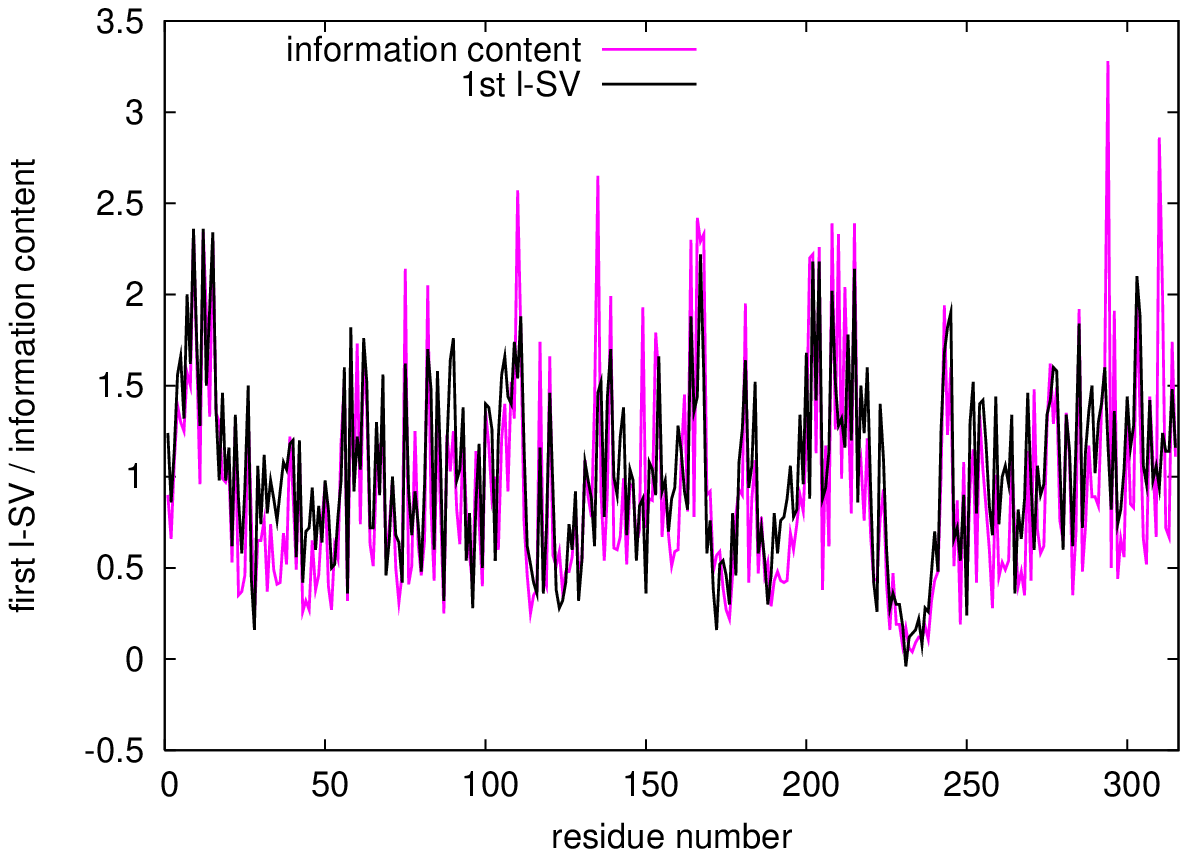} 
\end{center}

  \subsection*{Figure 6 - Example of the second left singular vector and contact numbers.}
Shown are the second l-SV and contact numbers of the PDB entry 1l2hA\cite{1L2H}. The 
values of the l-SV elements are shifted and scaled to match the contact 
numbers. The correlation coefficient is 0.71.
\begin{center}
 \includegraphics[width=8cm]{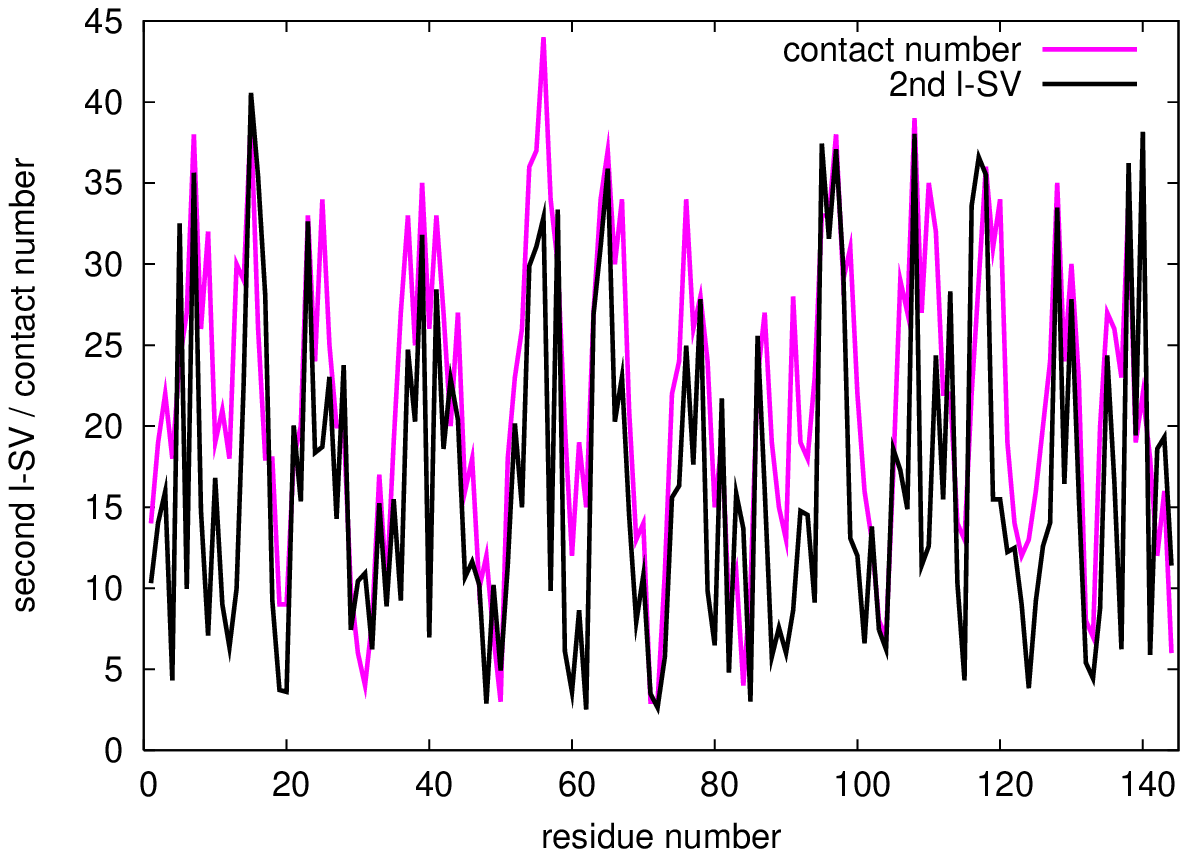} 
\end{center}

  \subsection*{Figure 7 - Decomposed PSSM scores of conserved sites in the globin family.}
The labels B10, C2, CD1, CD4, E7, F4, and F8 on the top of panels are site 
identifiers of the globin family defined by Bashford et al.\cite{BashfordETAL1987} (in the parentheses is the most conserved residue at each site). 
The PSSM is based on the PDB entry 3sdhA\cite{3SDH}.
\begin{center}
  \includegraphics[width=14cm]{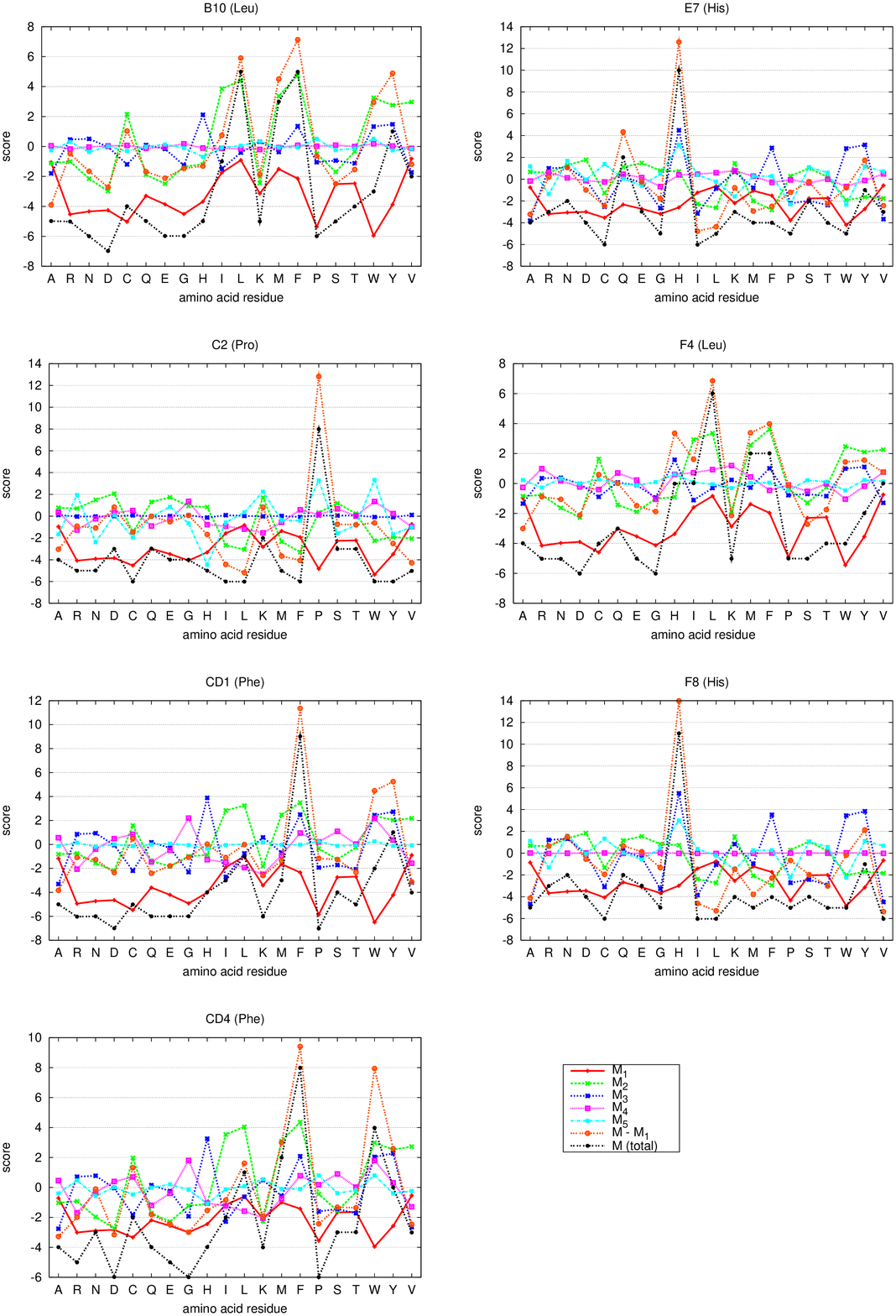}
\end{center}
%%%%%%%%%%%%%%%%%%%%%%%%%%%%%%%%%%%
%%                               %%
%% Tables                        %%
%%                               %%
%%%%%%%%%%%%%%%%%%%%%%%%%%%%%%%%%%%

%% Use of \listoftables is discouraged.
%%
\section*{Tables}
  \subsection*{Table 1 - AAindex entries mentioned in the text.}
  \mbox{
  \begin{tabular*}{\columnwidth}{@{\extracolsep{\fill}}lp{15em}l}
\hline
ID & Description & Reference \\\hline
AURR980119 & Normalized positional residue frequency at helix termini C"' & Aurora and Rose (1998)\cite{AuroraANDRose1998}\\
BASU050101 & Interactivity scale obtained from the contact matrix & Bastolla et al. (2005)\cite{BastollaETAL2005}\\
BASU050103 & Interactivity scale obtained by maximizing the mean of correlation coefficient over pairs of sequences sharing the TIM barrel fold & Bastolla et al. (2005)\cite{BastollaETAL2005}\\
BEGF750103 & Conformational parameter of beta-turn & Beghin and Dirkx (1975)\cite{BeghinANDDirkx1975}\\
BUNA790101 & alpha-NH chemical shifts & Bundi and Wuthrich (1979)\cite{BundiANDWuthrich1979}\\
CHAM830106 & The number of bonds in the longest chain & Charton and Charton (1983)\cite{ChartonANDCharton1983}\\
FAUJ880106 & STERIMOL maximum width of the side chain & Fauchere et al. (1988)\cite{FauchereETAL1988}\\
FUKS010106 & Interior composition of amino acids in intracellular proteins of mesophiles & Fukuchi and Nishikawa (2001)\cite{FukuchiANDNishikawa2001}\\
GRAR740102 & Polarity & Grantham (1974)\cite{Grantham1974}\\
JOND920102 & Relative mutability & Jones et al. (1992)\cite{JonesETAL1992b}\\
KLEP840101 & Net charge & Klein et al. (1984)\cite{KleinETAL1984}\\
KOEP990101 & Alpha-helix propensity derived from designed sequences & Koehl and Levitt (1999)\cite{KoehlANDLevitt1999b}\\
KYTJ820101 & Hydropathy index & Kyte and Doolittle (1982)\cite{KyteANDDoolittle1982}\\
LEVM760102 & Distance between C-alpha and centroid of side chain & Levitt (1976)\cite{Levitt1976}\\
LEVM760105 & Radius of gyration of side chain & Levitt (1976)\cite{Levitt1976}\\
MIYS990101 & Relative partition energies derived by the Bethe approximation & Miyazawa and Jernigan (1999)\cite{MiyazawaANDJernigan1999}\\
OOBM770105 & Short and medium range non-bonded energy per residue & Oobatake and Ooi (1977)\cite{OobatakeANDOoi1977}\\
SNEP660101 & Principal component I & Sneath (1966)\cite{Sneath1966}\\
SNEP660103 & Principal component III & Sneath (1966)\cite{Sneath1966}\\
SWER830101 & Optimal matching hydrophobicity & Sweet and Eisenberg (1983)\cite{SweetANDEisenberg1983}\\
\hline
  \end{tabular*}}
 
  \subsection*{Table 2 - Amino acid indices most correlated to the first right singular vectors [frequency (\%) in the parentheses].}
The description of each AAindex ID can be found at \texttt{http://www.genome.jp/dbget-bin/www\_bfind?aaindex}.
\mbox{
  \begin{tabular}{rll}\hline
rank & PDB & Pfam \\\hline
 1 & JOND920102 (10) & SNEP660101 (9) \\
 2 & FUKS010106 (7) & DESM900101 (7) \\
 3 & MCMT640101 (6) & KYTJ820101 (6) \\
 4 & MEEJ810101 (6) & WOLS870102 (6) \\
 5 & BEGF750103 (5) & JOND920102 (6) \\
 6 & KYTJ820101 (4) & FUKS010106 (5) \\
 7 & ROBB790101 (4) & BEGF750103 (3) \\
 8 & KIDA850101 (3) & CORJ870108 (3) \\
 9 & ROBB760108 (3) & LEVM780106 (2) \\
10 & MIYS990101 (3) & AURR980120 (2) \\
\hline
  \end{tabular}}
 
  \subsection*{Table 3 - Amino acid indices most correlated to the second right singular vectors [frequency (\%) in the parentheses].}
\mbox{
  \begin{tabular}{rll}\hline
rank & PDB & Pfam \\\hline
 1 & MIYS990101 (33) & MIYS990101 (54) \\
 2 & BASU050101 (12) & GRAR740102 (12) \\
 3 & BASU050103 (11) & BASU050103 (10) \\
 4 & GRAR740102 (8) & MIYS990102 (7) \\
 5 & SWER830101 (7) & BASU050101 (6) \\
 6 & MIYS990102 (6) & SWER830101 (1) \\
 7 & ZHOH040103 (2) & ZHOH040103 (1) \\
 8 & CORJ870102 (2) & MIYS990105 (1) \\
 9 & KYTJ820101 (2) & FAUJ830101 (1) \\
10 & FAUJ830101 (2) & CORJ870102 (1) \\
\hline
  \end{tabular}}
 
  \subsection*{Table 4 - Amino acid indices most correlated to the third right singular vectors [frequency (\%) in the parentheses].}
\mbox{
  \begin{tabular}{rll}\hline
rank & PDB & Pfam \\\hline
 1 & CHAM830106 (20) & CHAM830106 (15) \\
 2 & SNEP660103 (18) & SNEP660103 (15) \\
 3 & LEVM760102 (12) & LEVM760102 (12) \\
 4 & KOEP990101 (8) & LEVM760105 (8) \\
 5 & OOBM770105 (6) & WOLS870102 (5) \\
 6 & LEVM760105 (6) & FASG760101 (5) \\
 7 & MITS020101 (6) & KOEP990101 (4) \\
 8 & HUTJ700103 (2) & CHAM830104 (4) \\
 9 & RADA880103 (2) & HUTJ700103 (3) \\
10 & CHAM830105 (2) & OOBM770105 (3) \\
\hline
  \end{tabular}}

%%%%%%%%%%%%%%%%%%%%%%%%%%%%%%%%%%%
%%                               %%
%% Additional Files              %%
%%                               %%
%%%%%%%%%%%%%%%%%%%%%%%%%%%%%%%%%%%

%\section*{Additional Files}
%  \subsection*{Additional file 1 --- Sample additional file title}

\end{bmcformat}
\end{document}